\documentclass[12pt]{article}

\usepackage{amssymb}
\usepackage{amsbsy}
\usepackage{amsmath}
\usepackage{bm}

\newcommand\ba{\begin{equation}}
\newcommand\ea{\end{equation}}

\def\TEM{{T^{(NLED)}}}

\def\JeD{{\cal J}^D_e}
\def\starG{{\star G}}
\def\starF{{\star F}}
\def\JU1{{\cal J}^{U1}}
\def\L{{\cal L}}
\def\LL{\hat{\cal L}}
\def\Vt{{\tilde V}}
\def\ot{\otimes}
\def\nab{{\nabla}}
\def\calP{{\cal P}}
\def\bfe{{\bf e}}
\def\bfb{{\bf b}}

\def\bfx{\hat{\boldsymbol x}}
\def\bfy{\hat{\boldsymbol y}}

\def\wtilde{\widetilde}
\def\tmax{{\star\tau^{(LED)}_{\widetilde{d N}}}}

\def\Jhat{{ \JU1 - \star( dN \wedge \starF )  - \star( dL\wedge F  )  }}

\def\NLVED{{ non-linear vacuum electrodynamics }}
\begin{document}
\begin{titlepage}

\title{Charged Relativistic Fluids and Non-linear Electrodynamics }

\author{T. Dereli\footnote{tdereli@ku.edu.tr} \\ {\small Department of Physics, Ko\c{c} University, 34450 Istanbul, Turkey }\\
 \\ R.  W. Tucker\footnote{r.tucker@lancaster.ac.uk}\\
{\small  The Cockcroft Institute, Daresbury, UK} \\ {\small and}
\\{\small Department of Physics, Lancaster University, Lancaster, UK
}}

\date{\empty}

\maketitle

\begin{abstract}
{\small  \noindent The electromagnetic fields in Maxwell's  theory
satisfy linear equations in the classical vacuum. This is modified
in classical non-linear electrodynamic theories. To date there has
been little experimental  evidence that any of these modified
theories are tenable. However with the advent of high-intensity
lasers and powerful laboratory magnetic fields this situation may be
changing. We argue that an approach involving the self-consistent
relativistic motion of a smooth fluid-like distribution of matter
(composed of a large number of charged or neutral particles) in an
electromagnetic field  offers a viable theoretical framework in
which to explore the experimental consequences of non-linear
electrodynamics. We construct such a model based on the theory of
Born and Infeld and suggest that a simple laboratory experiment
involving the propagation of light in a static magnetic field could
be used to place bounds on the fundamental coupling in that theory.
Such a framework has many applications including a new description
of the motion of particles in  modern accelerators and plasmas as
well as phenomena in astrophysical contexts such as in the
environment of magnetars, quasars  and gamma-ray bursts. }
\end{abstract}

\bigskip

\noindent Classification Numbers:
 02.40.Hw
,\, 03.50.De
,\, 41.20.-q

\end{titlepage}

\section{Introduction}
\label{ch1}

Maxwell's theory of classical electromagnetic phenomena employs
linear partial differential equations to describe the behaviour of
fields in source-free regions of the vacuum. The extension of the
theory to fields in material media may involve non-linear
modifications arising from the complex interactions between
distributions of charge at a fundamental level. However, with the
advent of high-power lasers and high-field gradients in plasmas one
may be approaching regimes where the linear nature of Maxwell vacuum
electrodynamics breaks down with attendant implications for
electrodynamics in media. Certainly one expects vacuum polarization
induced by quantum processes to intrude when electric field
strengths approach  $1.3 \times 10^{16}$ V/cm. This is still some
orders of magnitude greater than current field intensities in pulsed
lasers so it is of interest to enquire whether classical effects of
non-linear vacuum electrodynamics \cite{BI}, \cite{plebanski} may
yield experimental signatures before the need to accommodate quantum
phenomena. The role of \NLVED at  a fundamental level may offer new
insights
 into the problem of classical radiation reaction on particles and high-intensity field-particle interactions in plasmas.

One difficulty in assessing the significance of \NLVED is in constructing a tractable generalisation of Maxwell's
theory that is amenable on some scale to experimental verification. In this note we suggest that an approach involving
the self-consistent relativistic motion of a smooth fluid-like distribution of matter (composed of a large number of
charged or neutral particles) in an electromagnetic field  offers a viable theoretical framework in which to explore
experimental consequences. Such a framework has many applications including the motion of particles in  modern accelerators
and plasmas as well as phenomena in astrophysical contexts such as in the environment of magnetars, quasars  and gamma-ray bursts.

\noindent  In the following it is assumed that the electromagnetic
field is a 2-form $F$ on space-time with a metric tensor field
\ba{g= -e^0\otimes e^0 + \sum_{k=1}^3 e^k\otimes e^k} \ea where
$\{e^a\}$ is a local co-frame with dual basis $\{ X_a \}$ for
$a=0,1,2,3$. Furthermore it is assumed that locally $F$ can be
expressed in terms of the 1-form $A$ by $F=d\,A$ and that
electrically charged matter interacts with the field via a $U(1)$
covariant interaction giving rise to a regular 4-current density
3-form $\JU1  $. Singular sources, such as point charges contribute
a singular distributional  current $\JeD$. The generalized Maxwell
system for the field $F$ is taken to be

\ba{ d\,F= 0} \quad ,\ea

\ba{ d\,\starG= \JeD + \JU1}\ea
where $\star$ denotes the Hodge map associated with $g$. The 2-form
$G$  is related to $F$ by a constitutive relation which for \NLVED
is non-linear. In this note it is assumed that such a relation is
local and takes the form \ba{\starG= Z_1(F,g)}\ea for some tensor
$Z_1$. Furthermore it is assumed that the 3-form $\JU1$ may depend
locally on $F, g$  and a unit time-like 4-vector field $V$
describing a smooth flow of matter on space-time: \ba{
\JU1=Z_2(F,V,g) }\quad . \ea
 In the following the $1-$form $\wtilde V$ is related directly to the vector field $V$ by the metric.
 It is  defined by the relation $\wtilde{V}(X)=i_X\wtilde V= g(V,X)$ for all vector fields  $V$ and $i_{X_a}  $ is abbreviated $i_a$.

In the absence of matter the constitutive tensor $Z_1$ can be
derived from an action of the form \ba{
S_1[A,g]=\int_M\,\LL(F,\nabla\,F,\cdots, g) \,\star1}\ea involving
some Lagrangian 0-form $ \LL(F,\nabla\,F,\cdots, g) $. If one
further restricts to Lagrangians of the form
$\LL(F,\nabla\,F,\cdots, g)=\L(X,Y,g)$ where
 $X=\star(F\wedge \starF)$ and $Y=\star(F\wedge F) $
then \ba{\starG= Z_1(F,g) = 2\star F \L_X + 2 F\L_Y}\ea where $\L_X=
\frac{\partial \L}  {\partial X}$ and $\L_Y=\frac{\partial \L} {\partial
Y}  $.  The vacuum stress-energy-momentum tensor $\TEM$ follows from
metric variations of $  S_1[A,g] $ as
 $\TEM= (\star \tau_a) \otimes e^a $
where \ba{ \tau_a =M\, \star e_a + N \,\tau_a^{(LED )}  } \ea with $
M= ( \L -X\L_X -Y \L_Y ) $,  $  N= 2 \L_X  $ and \ba{ \tau_a^{(LED
)} = \frac{1}{2} \left( i_aF \wedge \starF - i_a\starF \wedge F
\right) }\quad . \ea The dependence of the forms $\tau_a^{(LED )}$
on $F$ is the same as that in Maxwell's linear electrodynamics in
the vacuum.

\section{Charged Thermodynamic Fluids}
\label{ch2} Consider matter with proper mass-energy density $\rho$,
proper charge density $\rho_e$ and convective electric 4-current
$\JU1=\rho_e \star \wtilde V$.
Its equation of motion  is given by \ba{\nabla \cdot T_{(total)}=0
\label{divT}} \ea for some total stress-energy-momentum tensor
\ba{T_{(total)}= \TEM + T_{( fluid)} }\quad . \ea For a fluid
without dissipation but thermodynamic pressure $p$  it will be
assumed that \ba{ T_{( fluid)} = (\rho + p   ) \Vt \ot \Vt + p\, g
}\quad .\ea It follows immediately that \ba{ \widetilde{\nab \cdot
T_{( fluid)}} = V\,\nab \cdot ((\rho+p)V   ) + (\rho+p)\nab_V\,V
+\widetilde{d\,p} }\, .\ea
The divergence of $\TEM  $ is more complicated and it is convenient
to introduce some abbreviations. For any vector field $Q$  with
ortho-normal components $Q^a$  write  $\tau^{(LED )}_Q=  \tau^{(LED
)}_{a}\,Q^a $ so \ba{ \nab\cdot\TEM= d\,M +  \tmax - i_{\hat J} F
}\ea where \ba{\hat J=  \Jhat  }\ea with  $L= 2\L_Y$. Equation
(\ref{divT}) yields \ba{  d\,M +  \tmax - i_{\hat J} F V\,\nab \cdot
((\rho+p)V   ) + (\rho+p)\nab_V\,V +\widetilde{d\,p}=0
\label{master} }\ea  which upon contracting with $V$ gives the
tangential component continuity equation \ba{ (p+\rho)\, \nab\cdot
V= i_V\,d\,M - i_V\tmax - V(\rho) - i_V\,i_{\hat J} F  }\quad . \ea
Substituting this into (\ref{master}) yields, in terms of the
projection operator $\Pi_V= (1+ \Vt\,i_V  )$, the relativistic fluid
equation of motion \ba{ (\rho +p)\, \widetilde{\nab_V\,V}= \Pi_V\,
\calP }\ea where the total pressure 1-form
$$ \calP= i_{\hat J}\,F - d\,M -d\,p-\tmax \; . $$

The proper mass-energy density $\rho(\rho_m,p)$  can be expressed in
terms of the proper mass density  $\rho_m$ and the pressure $p$
given a specific internal energy function  $ {\cal E}\left(\rho_m,p
\right) $: \ba{ \rho( \rho_m,p )=\rho_m\left(  1 +   {\cal
E}(\rho_m,p ) \right)   }\ea The thermodynamic temperature $T$ and
entropy ${\cal S}$ of the fluid are defined via the relation \ba{T
\,d{\cal S} = d\,{\cal E} + p\,d\left(\frac{1}{\rho_m}\right)}\ea
which may be expressed in terms of $d\,\rho_m$ and $d\,p$.

\section{Born-Infeld Fluids}
\label{ch3}

 Born-Infeld \NLVED has much to recommend it \cite{boillat}, \cite{deser}. Aside from its historic significance it is thought to encapsulate
 aspects of effective string theory \cite{tseytlin}, \cite{gibbons} including electromagnetic duality covariance. Here it will be adopted in a gravity free environment and its salient features
 explored in the context of the relativistic fluid.
The Lagrangian takes the form \ba{ \L(X,Y,g)=
\frac{\epsilon_0}{\kappa^2} (1- \sqrt{\Delta(X,Y)})  }\ea with
$$\Delta(X,Y)=1 - \kappa^2\,X - \frac{\kappa^4}{4} \,Y^2$$
\ba{2\L_X= \frac{\epsilon_0}{\kappa^2\sqrt{\Delta}}  }\ea \ba{2\L_Y=
\frac{\epsilon_0\,Y}{2\sqrt{\Delta}}  }\ea and is governed by a new
constant of nature\footnote{The fundamental constant $\kappa$ has SI
dimensions $[\frac{Q \,T^2}{M\,L}]$ and $\epsilon_0$ is the
permittivity of free space.   } $\kappa$. It follows that the vacuum
constitutive relation is \ba{ \starG=
\frac{\epsilon_0}{\kappa^2\,\sqrt{\Delta}}\left( \starF +
\frac{\kappa^2}{2}Y\,F  \right) }\ea and the fluid system can be
rewritten as \ba{ (\rho + p)\, \widetilde{\nab_V\,V}= \Pi_V\left(
i_{\hat J} \,F + i_\eta\,F - d\,p - \xi \right)  }\ea where
$$ \xi \equiv d\,M + \tmax \; , $$
$$ \tilde\eta \equiv \star( d\,N \wedge \starF ) + \star( d\,L \wedge F )\label{master2} \; .$$
With $\JU1=\rho_e\,V$  one has $\Pi_V\, i_{\hat J}\, F = \rho_e\,
i_V F$ and (\ref{master2}) yields: \ba{(\rho + p)\,
\widetilde{\nab_V\,V}= \rho_e\,i_V\,F + \Pi_V\,(i_\eta\,F - d\,p
-\xi    )} \; . \ea

In Maxwell electrodynamics $$ {\cal L}=X  $$ and hence  $\eta=\xi=0$
and the system reduces to:
 $${ V \cdot V=-1,\quad d\,F=0,\quad d\,\star{G}=\rho_e \star \tilde{V}, \quad{G}= \epsilon_0\,F} \quad , $$
{$$ (\rho+p) \nabla_V\tilde{V}  = \rho_e\,F(V)  -d\,p \; , $$
$$ (\rho+p) \, \nabla\cdot V=-V(\rho) $$}
exhibiting flow under the Lorentz force $  \rho_e\,F(V) $  and
thermodynamic pressure gradients.

By contrast, in the Born-Infeld electrodynamics, even in the absence
of electrically charged matter couplings contributing to  $\hat J$
via the   $U(1)$ electric current $\JU1$, there is a non-zero
Born-Infeld electro-dynamic pressure $\Pi_V(\xi -i_\eta\,F)$
contributing to the total pressure on the fluid: \ba{(\rho+p)\,
\widetilde{\nab_V\,V}  = -\Pi_V(\xi -i_\eta\,F) - \Pi_V\, d\,p
\label{A1}}\quad . \ea This equation together with the above continuity
equation: \ba{(\rho+p)\,\nab\cdot V= -V(\rho) + i_V(\xi + i_\eta\,F
) \label{C1} }\ea and Born-Infeld field equations in the absence of
singularities \ba{ d\,F=0 \quad , \quad d\,\starG=0 \label{BI} }\ea
constitute the equations for a coupled $U(1)$-neutral, relativistic
Born-Infeld thermodynamic fluid \cite{dirac}, \cite{chruscinski}. If
one can neglect collisions and internal energy, one has a
$U(1)$-neutral cold, thermodynamically inert fluid (dust) satisfying
(\ref{BI}), (\ref{A1}) and  (\ref{C1}) with  $p=0, \rho=\rho_m$. The
remaining electrodynamic pressures may arise whenever $\eta$ and
$\xi$ are non-zero for fields $F$ such that $d\,X\neq 0$ and $d
\,Y\neq 0$.

Such pressures can arise from solutions in Born-Infeld
electrodynamics in Minkowski space-time with plane propagating waves
superposed with a uniform static magnetic field  in vacuo
\cite{aiello et al}. A particular case is a  magnetic field
transverse to the direction of propagation of a plane wave with an
arbitrary smooth longitudinal profile ${\cal E}(z)  $:
$$F= {\cal E}(z-v t) \, d(z-v t)  \wedge d x - B\, d y \wedge d z  $$
describing the electric and magnetic fields in an inertial frame:
$$ \bfe= -\frac{v}{c} {\cal E}(z-v t) {\bfx}, \quad \bfb= \frac{{\cal E}(z-v t)}{c} \bfy - \frac{B}{c} \bfx$$ where
$$ v=\frac{c}{\sqrt{1+ c^2\kappa^2 B^2}} \; . $$
 Thus the static magnetic field with magnitude $B$ slows down the propagating electromagnetic field with amplitude proportional to
 ${\cal E}$ to a phase speed $v<c$ {\it in vacuo}. Since this retardation is cumulative it may be amenable to experimental analysis
 with laboratory magnetic fields.
If one sets $\kappa\simeq  \frac{\epsilon_0\,r_0^2}{e}$ in terms of
the classical radius of the electron\footnote{$r_0=
\frac{e^2}{4\pi\epsilon_0\, m_e c^2}\simeq 2.8E-15  $m} then the
Born-Infeld electron model bounds $\kappa<10^{-22}$. The wave
transit time difference between  when the static field is switched
on  and off in a magnet region of length $L_0$ is
$$\tau=
\frac{L_0}{2}\kappa \vert B\vert .\label{pop}$$
So for $\kappa<10^{-22}$, $L$ in metres and
$\vert B \vert$ in Tesla
$$\tau<\frac{L}{2}\vert B\vert 10^{-6} \mbox{ps}$$
where $1$ps$=10^{-12}$ sec. This suggests that a terrestrial experiment could be used to place bounds on the coupling $\kappa$.

\section{Conclusions}
A general model of a charged fluid interacting with an  electromagnetic field whose vacuum properties are governed by a
Lagrangian generalizing Maxwell's theory has been devised. It exhibits new pressure gradients of a purely electrodynamic
origin in addition to those expected from Maxwell's theory. These forces may exist even when the fluid is electrically
neutral in the vacuum. The particular case of a Born-Infeld fluid has been chosen to illustrate the existence of these
forces when the fluid moves in a background static magnetic field on which a plane wave of arbitrary longitudinal profile
propagates. The properties of this wave offer  a means to bound the fundamental Born-Infeld coupling.
Once one has bounds on $\kappa$  it is proposed that the framework above offers a new and intriguing avenue to explore
the effects of \NLVED in high field regimes that may become accessible to observation before the breakdown of classical electrodynamics.

\label{ch4}

\newpage

\section*{Acknowledgment}
RWT is grateful to colleagues at the Cockcroft Institute for
valuable discussions, to the EPSRC  for a Springboard Fellowship
 for financial support  for this
research which is part of the Alpha-X collaboration. We thank
Ko\c{c} University  for its hospitality where part of this research
is carried out and the Turkish Academy of Sciences (TUBA)  for a
travel grant.




\end{document}